\documentclass[a4paper,fleqn,usenatbib]{mnras}
\usepackage[T1]{fontenc}
\usepackage{ae,aecompl}

\usepackage{graphicx}	% Including figure files
\usepackage{amsmath}	% Advanced maths commands
\usepackage{amssymb}	% Extra maths symbols
\newcommand{\kms}{km s$^{-1}$}

\title[Unusual line profiles of SN~2006gy ]
{Explaining unusual line profiles of SN~2006gy}

\author[N. N. Chugai]{
Nikolai N. Chugai,$^{1}$\thanks{E-mail: nchugai@inasan.ru}
\\
Institute of Astronomy Russian Academy of Science, Moscow, Russia\\
}

\date{Accepted XXX. Received YYY; in original form ZZZ}

\pubyear{2016}

\begin{document}
\label{firstpage}
\pagerange{\pageref{firstpage}--\pageref{lastpage}}
\maketitle

% Abstract of the paper
\begin{abstract}
Origin of enigmatic line profiles of extremely luminous type IIn
supernova SN~2006gy on day 96 is explored. Among conceivable possibilities the 
most preferred is the model that suggests holes in the 
optically thick cool dense shell (CDS). The line radiation emitted 
at the inner side of the opaque CDS escapes through the holes thus producing 
unusual line profile with the emission shifted redward.
The holes could emerge as a result of the vigorous Rayleigh-Taylor instability 
leading to the CDS fragmentation. The model light curve with the CDS 
fragmentation is shown to be consistent with the SN~2006gy bolometric light curve.

\end{abstract}

% Select between one and six entries from the list of approved keywords.
% Don't make up new ones.

%%%%%%%%%%%%%%%%%%%%%%%%%%%%%%%%%%%%%%%%%%%%%%%%%%

%%%%%%%%%%%%%%%%% BODY OF PAPER %%%%%%%%%%%%%%%%%%

\section{Introduction} 
\label{sec:intro}

The supernova SN~2006gy is an exceptional case of strongly interacting 
type IIn supernovae (SNe~IIn) because of the tremendous luminosity 
($\sim2\times10^{44}$ erg s$^{-1}$) and the radiated energy  
($\sim2\times10^{51}$ erg)  \citep{Smith2007,Ofek2007,Smith2008,Smith2010}.
To account for this phenomenon  the ejecta with the energy 
$\gtrsim3\times10^{51}$ erg must 
collide with a massive circumstellar (CS) envelope  
($\sim10~M_{\odot}$), residing at the radius of $\sim3\times10^{15}$ cm 
\citep{SMC2007}. Unlike more common 
SNe~IIn, the radiation generated by the shock wave in the SN~2006gy is strongly  
trapped for about two months due to a large optical depth \citep{SMC2007}. 
Light curve modelling \citet{Moriya2013} based on the radiation hydrodynamics 
supports the picture of vigorous CS interaction and authors conclude 
that in this case SN ejecta of $\lesssim15~M_{\odot}$ 
and kinetic energy of $\gtrsim4\times10^{51}$ erg collides with the 
$\approx15~M_{\odot}$ of CS matter (CSM). 
The origin of the tremendous mass loss by the pre-SN and the very explosion 
mechanism are not fully understood. A thermonuclear explosion initiated by 
the pulsating pair-instability of massive star ($\approx100~M_{\odot}$) 
so far is the only viable scenario \citep{Woosley2007}.

Despite a general consensus on the vigorous CS interaction as a cause 
for the SN~2006gy phenomenon, some crucial issues of the 
spectrum formation are not understood.
The excellent spectral data and their comprehensive analysis 
\citep{Smith2010} reveal narrow and broad lines on the continuum which 
at first glance look like what one expects in SN~IIn. 
Yet \citet{Smith2010} come up with an interesting conclusion that 
broad line profiles, e.g.,  H$\alpha$ and 
\ion{Fe}{ii} 5018\,\AA\  at the age $t\gtrsim90$ d are highly unusual.
While possible line-forming are discussed by \citet{Smith2010}, 
the issue, what actually make the profiles odd, remains enigmatic.

I address the issue of the unusual broad lines in SN~2006gy in an attempt to 
find at least a qualitative explanation of the phenomenon.
The line profile depends primarily on the kinematics of 
line-forming regions, geometry of continuum-emitting gas and 
its position relative to the line-forming regions. 
Different possibilities will be explored using Monte Carlo simulations 
and most probable interpretation 
will be proposed (section \ref{sec:model}).
I will check then, whether constraints from the sensible line profile model 
are consistent with the bolometric light curve (section \ref{sec:lcurve}). 
The results are summarized and discussed in the conclusion section. 

%=============================================
% Example figure
\begin{figure}
    \includegraphics[width=0.9\columnwidth]{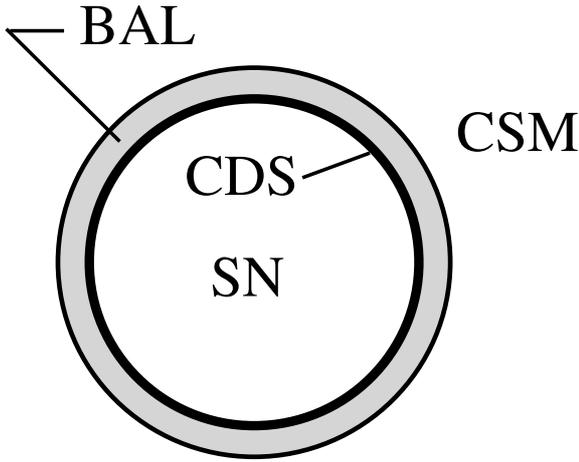}
    \caption{Sketch showing structure elements contributing to 
    line profile formation in the model A. Unshocked ejecta (SN) are 
    responsible for the bulk of a broad line emission; the partially transparent 
    CDS produces continuum; the BAL layer is responsible for the broad line 
    absorption and scattered emission; the unshocked CSM produces narrow lines.}
    \label{fig:cart}
\end{figure}

%=============================================

\section{Spectral modeling}
\label{sec:model}

\subsection{Preview}
\label{sec:preview}

SN~2006gy ejecta interact with a dense CSM in a strongly 
radiative regime which suggests that a cool dense shell (CDS) forms 
between the forward and reverse shocks (FS, RS). The CDS seems to be opaque 
for about 90 d and expands at $4000-5000$ \kms\ \citep{Smith2010}.
The opaque geometrically thin and spherically symmetric 
CDS with the attached FS prevents the formation of pronounced 
broad absorption lines of low-ionization species. That is why we see 
in early SN~2006gy only smooth continuum originated from the CDS 
with the CS emission lines 
broadened by Thomson scattering \citep{Smith2010} likewise, e.g., 
in SN~1998S \citep{Chugai2001}. 
Between days 71 and 93 the SN~2006gy spectrum 
\citep[cf.][]{Smith2010} changes dramatically: relatively smooth continuum 
becomes bumpy and new emission lines suddenly emerge with growing luminosity,
e.g., \ion{Ca}{ii} 
infrared triplet. This indicates that the opaque CDS becomes partially 
transparent at that epoch and line emission from the interior become visible.
It is about this age that the \ion{Fe}{ii} 5018\,\AA\ 
and H$\alpha$ lines acquire unusual profiles \citep{Smith2010} (cf. also 
figure~\ref{fig:sp1} below). The peculiar feature of these lines, most 
apparent in \ion{Fe}{ii} 5018\,\AA, is a red 
emission component flanked by a sharp blue cut-off 
at about zero radial velocity. 

\citet{Smith2010} conjecture that peculiar line profiles are due to 
a blue absorption with the monotonically increasing depth 
towards zero radial velocity. The absorbing gas lies presumably 
between the FS and the CDS. This picture is called here the model A (or case A). 
Indeed, a cool absorbing gas can form a layer between the CDS and CSM 
due to the Rayleigh-Taylor instability (RTI) of the CDS 
and subsequent mixing of RT spikes in the FS 
\citep{Chevalier1982a,CheBlo1995,BloEllis2001}. 
The major structure components involved in the 
broad line formation in the case A on day 96 then  
include the partially transparent CDS responsible for the 
continuum, the SN ejecta that produces broad emission lines partially 
attenuated by the CDS, and the outer broad absorption line (BAL) layer 
attached to the CDS  (Fig. \ref{fig:cart}).
The expected kinematics of the cool material in this case is the 
flow with almost constant velocity along the radius. The line profile in 
this case is well known and, anticipating simulations, unlikely able 
to fit the observed profiles. 
However, with some modification the model A still may be viable.

A partial transparency of the CDS in the considered epoch and  
especially a redshift of the broad emission lines by $\approx 1000$ \kms\ 
which is apparent in \ion{Fe}{ii} 5018\,\AA\ and \ion{Ca}{ii} 8662\,\AA\ 
\citep[cf.][]{Smith2010} prompts an alternative possibility (model B). It 
suggests that a significant fraction of the line emission originates 
at the inner side of the opaque CDS 
and this emission escapes through the holes in the CDS possibly after 
the Thomson scattering 
and absorption in the unshocked ejecta. Note this model does not include 
BAL layer. The cartoon  for the model B would be almost  
identical to Fig. \ref{fig:cart} but the BAL layer. 

Simulations of line profiles below are based on the Monte Carlo technique. 
Note that the models do not include narrow lines originated from the CSM.

\subsection{Model A}
\label{sec:mod_a} 

\begin{figure}
    \includegraphics[width=\columnwidth]{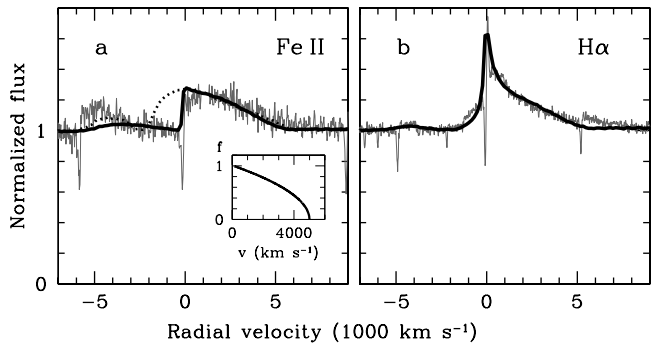}
    \caption{\ion{Fe}{ii} 5018\,\AA\ and H$\alpha$ line profiles in SN~2006gy 
    on day 96.
     Left panel: observed \ion{Fe}{ii} profile (grey thin line) compared to 
     the model A with constant velocity in the BAL layer (dotted line); the 
     model is strongly disfavoured by data. 
     The modified model A with the random velocities of line-absorbing gas 
     (thick line) fits data better. The inset shows the velocity distribution 
     in the BAL layer.
     Right panel: observed H$\alpha$ profile (grey thin line) with the overplotted 
     model with the random velocity distribution in the BAL layer.
     The model is the same as for \ion{Fe}{ii} except for the larger 
     line scattering albedo in the case of H$\alpha$.}
    \label{fig:sp1}
\end{figure}

The fiducial model A includes the partially transparent 
continuum-emitting CDS with the 
effective optical depth $\tau_s = 1$ and velocity $v_s = 5000$ \kms, 
the unshocked SN ejecta ($r < r_1 = 1$) with the 
terminal velocity $v_{sn} = 6000$ \kms, and the BAL layer 
($r_1 < r < r_2 = 1.2$) with the constant velocity $v = v_s = 5000$ \kms.
The line emissivity in the ejecta is adjusted to produce a sensible fit.
The Sobolev optical depth is set to be constant in the line-absorbing layer which
is not critical. 
The absence of the prominent emission component in the 
narrow component of \ion{Fe}{ii} 5018\,\AA\ line \citep[cf.][]{Smith2010} 
suggests that the scattering in this line is non-conservative, probably 
because of a photon destruction 
in a Rosseland cycle. This effect is taken into account assuming that 
the effective scattering albedo $\omega < 1$. 
The albedo stands for the ratio of the escaped  
radiation to the incident radiation and is
defined through the Sobolev escape 
probability ($\beta$) and the photon destruction probability ($\epsilon$) 
as $\omega = \beta/[\beta+(1-\beta)\epsilon]$.
The shown model profile (Fig. \ref{fig:sp1}a) 
is calculated for  $\omega = 0.5$. In the case $\omega = 0.1$ the profile 
differs insignificantly and therefore is not plotted. The observed line profile, 
as expected, disfavours the fiducial model A with the constant velocity in the BAL 
layer.

The model A can be modified to produce more acceptable fit.
To this end let us assume that the BAL layer 
contains an additional component related to the shocked 
CS clumps (if any) in the FS. The fragmentation and acceleration of shocked 
CS clumps can produce velocity spectrum of the cool gas in a broad range
between the cloud shock and the forward shock speeds \citep{Klein1994}. 
We assume that the velocity distribution $f(v)$ of the 
cool gas in the BAL layer is uniform along the radius.
The velocity presumably is dominated by the radial component 
$200 < v < 5000$ \kms\ 
 with an additional isotropic random component $0 < u < 0.2v$. 
We show the case of $\omega = 0.1$ that is preferred 
compared to $\omega = 0.5$. The adopted velocity distribution is 
shown in the inset. The modified model A provides an 
overall description of the \ion{Fe}{ii} 5018\,\AA\ line
 (Fig. \ref{fig:sp1}a) although the redshift of the emission is not reproduced. 
The same model, but $\omega = 1$, is applied to the H$\alpha$.  
The blue emission in this model is 
due to the resonant scattering in the BAL layer.
The fit in the low velocity region is not good enough; 
parameter variations does not remove the disparity.
In fact, a more serious problem with this model is that the SN~2006gy 
does not show an apparent signature of the CSM clumpiness. Indeed, 
the spectra do not show evidence for the 
intermediate component which usually is associated with shocked CS clumps. 

\subsection{Model B}
\label{sec:mod_b} 

The alternative case B includes three major components: 
an opaque CDS with transparent holes, unshocked SN ejecta, 
and a thin line-emitting layer at the inner side of the CDS.
The brightness of the outer and inner surface of the CDS in the continuum 
is assumed to be similar and isotropic. The intensity of the line 
radiation emitted by the inner side of the CDS is also assumed to 
be isotropic. The unshocked ejecta has large Sobolev 
optical depth and can be a source of the net line emission. 
The CDS velocity is $v_s = 5000$ \kms\ and the SN ejecta 
terminal velocity is adopted to be $v_{sn} = 7000$ \kms, the value 
consistent with the   \ion{Fe}{ii} 4924 \AA\ and 5018 \AA\ lines and 
the light curve model, as 
we see below. The ejecta Thomson optical depth ($\tau_{t}$),
the absorption optical depths ($\tau_a$), and hole covering factor ($h$) are 
free parameters. 

\begin{figure}
    \includegraphics[width=\columnwidth]{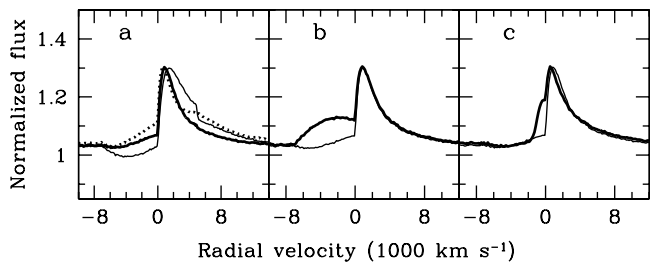}
    \caption{Illustrative models for case B (see Table 1). 
    Left panel shows the model m1 (thin line), model m2 (thick line) with larger 
    absorption, and model m3 (dotted line) with 
    larger Thomson optical depth. Middle panel compares model m2 (thin line) 
    and the  model m4 with albedo $\omega = 1$. Right panel shows model m2 
    (thin line) and m5 that includes the line emission from ejecta.}
    \label{fig:sp2}
\end{figure}

The effect of parameter variations is shown in Fig. \ref{fig:sp2} with 
major parameters (Thomson optical depth, absorption optical depth, albedo,
and fraction of net emission from the unshocked ejecta) given in Table 1.
Results are not sensitive to the hole covering factor 
and $h = 0.3$ is adopted; this formally corresponds to the CDS 
 effective optical depth of unity.
The line profile is rather sensitive to the optical 
depth both $\tau_{t}$ and $\tau_a$ (Fig. \ref{fig:sp2}a). 
The effect of the line scattering albedo is also strong   
(Fig. \ref{fig:sp2}b) and thus can be easily constrained. The contribution 
of the line net emission from the SN ejecta with the fraction 
$f_{ej} = 0.4$ results in the significant modification of the line profile 
(Fig.~\ref{fig:sp2}c); this behaviour is 
crucial for the \ion{Fe}{ii} line and the H$\alpha$ modelling. 
%================================================================
\begin{table}
 \caption{Parameters of illustrative models in case B.}
 \label{tab:example}
 \begin{tabular}{lcccc}
  \hline
  Model & $\tau_t$ & $\tau_a$ & $\omega$ & $f_{ej}$\\
  \hline
  m1 &    1 &  1 &  0  &  0 \\
  m2 &    1 &  2 &  0  &  0 \\
  m3 &    2 &  2 &  0  &  0 \\
  m4 &    1 &  2 &  1  &  0 \\
  m5 &    1 &  2 &  0  &  0.4 \\
  \hline
 \end{tabular}
\end{table}
%=================================================================

\begin{figure}
    \includegraphics[width=\columnwidth]{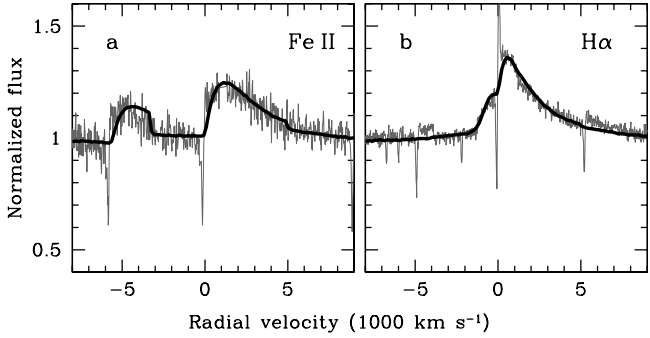}
    \caption{\ion{Fe}{ii} and H$\alpha$ lines     \
    in SN~2006gy on day 96 (gray thin line) with the overplotted model B. 
    Left panel shows \ion{Fe}{ii} 4924 \AA\ and 5018 \AA\ lines, and right panel 
    shows H$\alpha$.     
    The emission excess in the range of radial velocities 5000-7000 \kms\  
    is due to \ion{He}{i} 6678 \AA\ that is absent in the model.    
    }
    \label{fig:sp3}
\end{figure}

The model B is applied first to both \ion{Fe}{ii} lines of 42 multiplet, 
4924 \AA\ and 5018 \AA. These emissions in astrophysical objects 
have comparable fluxes with the 5018 \AA\ line to be somewhat stronger. 
In the $\eta$ Car spectrum \citep{Zethson2012} the line intensity ratio 
$I(4924)/I(5018) = 0.68$; this value is used here. In SN~2006gy on day 96 the 
\ion{Fe}{ii} 4924 \AA\ emission is significantly weaker and more narrow than 
5018 \AA. This is due to the absorption in the unshocked ejecta 
of 4924 \AA\ redshifted photons by the 5018 \AA\ line, the effect included in 
the model (Fig.~\ref{fig:sp3}). The red cut-off of the 4924 \AA\ 
line emission is consistent with the adopted terminal velocity of the 
unshocked ejecta (7000 \kms), although noise does not permit to constrain 
this velocity better than $\pm1000$ \kms. The \ion{Fe}{ii} net emission 
in the presented model fully originates from the inner side of the CDS. 
The line scattering albedo is constrained as $\omega < 0.2$; here 
$\omega = 0.1$ is used. Other parameters are $\tau_t = 0.7$,
 $\tau_a = 1.5$ (the uncertainty is 10-15\%).
The model describes the 4924 \AA\ line not as good as the 5018 \AA\ 
possibly because of an interference of the former with 
the H$\beta$ emission; this effect is not included in the model. 
The H$\alpha$ model, unlike the \ion{Fe}{ii} case, includes the net emission 
from unshocked ejecta with 24\% of the total line emission rate. 
This permits us to describe the overall line shape, particularly. 
in the range of $\pm1500$ \kms. Another difference is 
the larger absorption in the ejecta, $\tau_a = 2.2$ compared to 
1.5 for \ion{Fe}{ii} lines. 
Generally this is expected since the absorption in the optical band 
increases from blue to red (Paschen continuum, H$^-$).

Summing up, both \ion{Fe}{ii} 5018 \AA\ line and H$\alpha$ are 
described by the model B more successfully compared to the model A. Note, 
the model B easily reproduces the redshift of emission maximum 
of $\approx 1000$ \kms\ in agreement with the observed redshift in 
\ion{Fe}{ii} 5018 \AA\ and \ion{Ca}{ii} 8662 \AA\ emissions 
\citep[cf.][]{Smith2010}, whereas this effect is absent in the model A.

%================================================================
% Example figure
%=============================================================

%================================================================
% Example figure
%=============================================================

%================================================================
% Example figure
%=============================================================

\section{CDS fragmentation and light curve}
\label{sec:lcurve} 

The proposed interpretation of the unusual line profiles in both models 
suggests that the CDS on day 96 is partially transparent, $\tau_s \sim 1$. 
This raises a question, whether the proposed CDS fragmentation  
is consistent with the bolometric light curve. To explore this issue 
the CS interaction model should be able to compute the bolometric light curve 
with the radiation trapping and the CDS fragmentation.
This can be done in the framework of the model 
applied recently to the type IIn-P supernova SN~2011ht \citep{Chugai2016}.
Let us recap briefly the essential points of this model.

The model is based on a thin shell approximation \citet{Giuliani1982}.
The radiation trapping effects are described in a picture of 
a radiation bubble confined in an optically thick spherical shell. 
The bubble energy $E_r$ is determined by the energy generation in shock waves, 
the work spent by the radiation on the shell expansion,
and the radiation escape. The luminosity is defined as 
$L = E_r/t_d$ with the 
diffusion time $t_d = \xi (\tau + 1)r_s/c$, where  
$\tau$ is the total optical depth, $r_s$ is the shell radius, 
$c$ is the speed of light, and  $\xi$ is a fudge factor adopted to be 
$\xi = 0.5$; this value corresponds to the diffusion time in the problem 
of a central point source in a homogeneous sphere  \citep{Sunyaev1980}.
The total optical depth of the ejecta, CDS and CSM is defined by the 
Rosseland opacity \citep{Alexander1975}.  
The temperature distribution and the optical depth are recovered iteratively 
 on the bases of the Eddington approximation, 
$T^4 = (3/4)T_{eff}^4(\tau+2/3)$ where $T_{eff}$ is the effective temperature.
Usually 4-5 iteration is enough to attain the accuracy of one percent.
The thin shell density is set to be $\rho_s = 7\rho_{cs}$ ($\rho_{cs}$ is a preshock 
CS density) in line with the 
density jump in the strong radiation-dominated shock.
Unshocked supernova ejecta is assumed to expand homologously, 
($v=r/t$) with the density  $\rho \propto v^{-1}$ in inner layers
and $\propto v^{-8}$ in outer layers. The density  in the 
CS envelope expanding at the velocity 150 \kms\ is set by the broken power 
law $\rho\propto r^{k}$ with 
the density maximum at $r_1$, $k = k_1 > 1$ for $r < r_1$, $k = k_2 < 0$ 
for $r_1 < r < r_2$, and $k = k_3 < 0$ for $r > r_2$.

\begin{figure}
\includegraphics[width=\columnwidth]{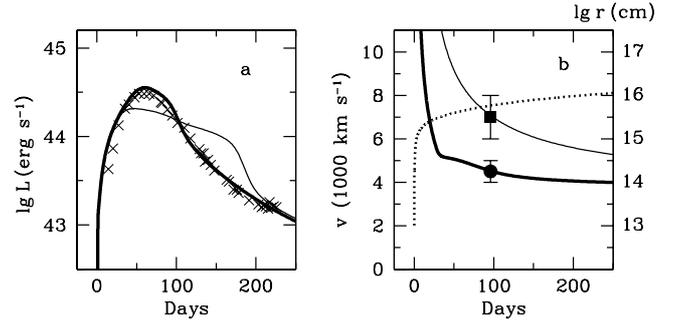}
\caption{SN 2006gy light curve, velocity and radius. 
Left panel: the observed bolometric light curve (crosses) with 
overplotted models, viz. the model with fragmented CDS (thick line) and 
intact CDS (thin line). Right panel: the model CDS velocity (thick solid line),
the terminal velocity of unshocked ejecta (thin solid line), and the CDS 
radius (dotted line). The points show the estimated velocity of the CDS (circle) 
and the terminal velocity of the unshocked SN ejecta (square) on day 96.}
\label{fig:blc}
\end{figure}

Keeping in mind that CDS is unstable with respect to RTI two cases are 
considered: intact and fragmented CDS. 
The CDS fragmentation is implemented via the effective optical depth 
of a clumpy medium 
\begin{equation}
\tau=\tau_{oc}\left[1 - \exp{\left(-\tau_0/\tau_{oc}\right)}\right]\,,
\label{eq:tau}
\end{equation}
\citep[cf.][]{Chugai2016}, 
where $\tau_0$ is the optical depth of the intact CDS and $\tau_{oc}$ 
is the occultation optical depth, i.e., the average number of the CDS 
fragments (clumps) along the radius. The evolution of this value is 
set as 
\begin{equation}
\tau_{oc}=\tau_{oc,0}\left[1+\left(\frac{t}{t_f}\right)^4\right]^{-1}\,,
\label{eq:tauoc}
\end{equation}
where for the optimal model $\tau_{oc,0}=108$, and $t_f=23$ d is the 
epoch of the fragmentation turn-on.
Anticipating results, the right panel of Fig.~\ref{fig:blc} 
suggests the vigorous CDS deceleration favouring RTI takes 
place before day 30.
The system of equations of motion, energy, and mass conservation 
is solved by the 4-th order Runge-Kutta. In each computed case the energy is 
conserved within  2\%.

%===============================================

%======================================================
The overall properties of the 
light maximum are found to be an outcome of a collision between  
$9~M_{\odot}$ SN ejecta with the kinetic energy of $5.3\times10^{51}$ erg 
and CSM of  $9~M_{\odot}$, in general agreement with results of \citet{Moriya2013}.
Two cases are plotted: the intact and fragmented CDS (Figure~\ref{fig:blc}).
The CSM is set by the radii $r_1 = 3\times10^{15}$ cm, 
$r_2 = 6\times10^{15}$ cm, $r_3=2\times10^{16}$ cm, and power law 
indices  $k_1 = 3$, $k_2=-3$, and $k_3=-5$.
The model with the intact CDS poorly describes the observed bolometric light curve 
\citep{Smith2010}. Note that the "shoulder" at the age of 100-140 d 
is a specific feature not only 
to our model; a similar shoulder is present in the light curve models 
of SN~2006gy computed in the framework of the radiation hydrodynamics 
\citep[figure 5]{Moriya2013}. 
In the model with the fragmentation the shoulder disappears and the 
overall fit gets better (Fig.~\ref{fig:blc}).
On day 96 the effective optical depth of the intact CDS is 
$\approx15$, whereas the optical depth of the fragmented CDS is $\approx1.3$, 
comparable to the CDS optical depth of the spectral model. 
Remarkably, the model CDS velocity on day 96 is $4500$ \kms, close 
to the observational estimate  $4000-5000$ \kms\ \citep{Smith2010}, and the 
model terminal ejecta velocity is also consistent with the estimate 
$v_{sn}\approx 7000$ \kms\ recovered from the \ion{Fe}{ii} line modelling.

To summarize, the CDS fragmentation invoked for the spectral model 
does not contradict to the bolometric light curve; moreover, the model 
with the fragmentation seems to be preferred. 

\section{Conclusions}
\label{sec:concl}

The study has been aimed at the problem of the unusual line profile  
observed in the spectra of SN~2006gy after about day 90 
\citep{Smith2010}. Two major conceivable models with the external BAL 
layer  (model A) and without it (model B) have been explored and 
a more successful description of \ion{Fe}{ii} and H$\alpha$ lines is 
found in the model B. It includes 
an opaque spherical CDS with transparent holes, 
a line-emitting layer at the inner side of the opaque CDS, and  unshocked ejecta 
that can scatter the line radiation and be a source of a net line emission as well. 
The success of rather simple model B indicates that 
essential physics is caught. Remarkably, the conjecture of the 
semi-transparent CDS at the age of 96 d turns out consistent with the 
bolometric light curve.

Yet the model B looks rather unconventional and poses some questions.
The most critical point is the formation of the "punched" CDS. 
Although the Rayleigh-Taylor instability of the CDS is unavoidable, 
it is unclear whether the CDS fragmentation is able to produce a punched CDS 
with transparent holes.
On the other hand, the fact that after about day 80 the luminosity of H$\alpha$ and 
\ion{Ca}{ii} IR triplet rapidly increases, despite bolometric luminosity 
does not show a coeval change, suggests that the CDS 
actually becomes partially transparent at this age. 
Suppose, however, that holes are created, the next question then arises, 
whether the line emission rate is significantly larger at the inner side of the 
opaque CDS than at the outer side. To
answer these questions direct numerical simulations are needed with 
the realistic high resolution 3D radiation hydrodynamics and 
3D multi-group radiation transfer, which is currently beyond reach. 
For the time being, therefore, we are left with the situation when the 
unusual phenomenon 
is explained in the framework of a somewhat odd model.
An encouraging aspect, however, is that the proposed explanation of the unusual 
line profiles 
in SN~2006gy can be considered as a starting point for the realistic look 
(beyond 1D) at the density and thermal structure of line-forming regions 
in SN~2006gy.

\section*{Acknowledgements}

I thank Nathan Smith for kindly sending the spectrum of SN~2006gy. 
This work is supported by RNF grant 16-12-10519.

%%%%%%%%%%%%%%%%%%%%%%%%%%%%%%%%%%%%%%%%%%%%%%%%%%

%%%%%%%%%%%%%%%%%%%% REFERENCES %%%%%%%%%%%%%%%%%%

% The best way to enter references is to use BibTeX:
% if your bibtex file is called example.bib
\bibliographystyle{mnras}
\bibliography{ref06gy.bib}

% Alternatively you could enter them by hand, like this:
% This method is tedious and prone to error if you have lots of references
%\begin{thebibliography}{99}
%\bibitem[\protect\citeauthoryear{Author}{2012}]{Author2012}
%Author A.~N., 2013, Journal of Improbable Astronomy, 1, 1
%\bibitem[\protect\citeauthoryear{Others}{2013}]{Others2013}
%Others S., 2012, Journal of Interesting Stuff, 17, 198
%\end{thebibliography}

%%%%%%%%%%%%%%%%%%%%%%%%%%%%%%%%%%%%%%%%%%%%%%%%%%
% Don't change these lines
\bsp	% typesetting comment
\label{lastpage}
\end{document}